\documentclass[11pt]{article}
\usepackage{latexsym,amssymb,amsmath,amsthm,a4wide,setspace,enumerate,color,graphicx,bbm,bm,url,dsfont}
\usepackage{subfiles} 
\usepackage[margin=3cm]{geometry}
\usepackage{natbib}
\usepackage{cmbright}
\usepackage{xcolor}
\usepackage{array}
\usepackage{xcolor-solarized}
\pagecolor{solarized-base2}
\color{solarized-base02}
\usepackage{changepage}
\usepackage{booktabs}
\usepackage{pgf}
\usepackage{graphicx}
\usepackage{epsfig}

\usepackage{booktabs,caption}
\usepackage[flushleft]{threeparttable}
\usepackage{mwe}
\usepackage{tablefootnote}
\usepackage{xfrac}
  
\usepackage{graphicx}
  
\usepackage{authblk}
\author[1,2]{Anders Huitfeldt}

\affil[1]{\small Department of Mathematics, Ecole Polytechnique F\'ed\'erale de Lausanne (EPFL), Switzerland}
\affil[2]{\small Centre for Evidence-Based Medicine Odense (CEBMO) and Cochrane Denmark, Department of Clinical Research, University of Southern Denmark, Odense, Denmark}

\usepackage[english]{babel}
\usepackage[utf8]{inputenc}
\usepackage{csquotes}

\usepackage{chngcntr}
\usepackage{apptools}
\usepackage{amsthm}
\usepackage{multirow}

\theoremstyle{definition}

\newtheorem*{remark*}{Remark}

\newtheoremstyle{break}
  {\topsep}{0}%
  {\bfseries}{}%
  {\newline}{}%
\theoremstyle{break}
\newtheorem*{remarks*}{Remarks}
\newtheorem*{examples*}{Examples}

\makeatletter
\def\munderbar#1{\underline{\sbox\tw@{$#1$}\dp\tw@\z@\box\tw@}}
\makeatother

\usepackage{afterpage}
\usepackage{caption}

\usepackage{amsmath}
\usepackage{amsfonts}

\begin{document}
\title{\huge Mindel C. Sheps: Counted, Dead or Alive}
\date{\today}
\maketitle
\sloppy
\thispagestyle{empty}\normalsize

\newpage
 
\section{Introduction}

When deciding whether to recommend that their patients initiate a medical treatment (such as the administration of drug $A$ which may affect outcome $Y$), many clinicians prefer to think about the expected consequences in terms of the relative risk, that is, a multiplicative parameter that relates the risk that the patients would experience the outcome if treated (which we will denote $Pr[Y^{a=1}=1]$), to the risk that they would experience the outcome if untreated ($Pr[Y^{a=0}=1]$). For example, the Cochrane Handbook\citep{deeks_chapter_2020}  and the GRADE guidelines\citep{guyatt_grade_2013} both explicitly instruct clinicians to combine the patient-specific baseline risk with the relative risk in order to produce an estimate of a patient's risk under treatment.

While many epidemiologists take a perspective on effect measures that is consistent with the clinicians' view, statisticians have tended to be uneasy about data analyses that rely on relative risk models, in part because such analyses may lead to predictions outside the range of valid probabilities, and in part because the models may lead to different predictions depending on whether they are based on the probability of the outcome event (say, death), or the probability of the complement of the outcome event (say, being alive)\citep{cox_analysis_1972}. Stated more formally, the relative risk model is not closed on the interval [0,1] and not invariant to choice of reference level for the outcome. These limitations of relative risk models have contributed to a longstanding difference in opinion between thought leaders from the three academic traditions, which frequently re-erupts in the scientific literature\citep{sackett_down_1996, walter_odds_1998}, on social media and on academic discussions forums.  

A simple solution to these limitations has been known since the 1950s but is not widely known or utilized. This solution was first described by Canadian physician and biostatistician Mindel Cherniack Sheps (1913-1973) in \textit{Shall we count the living or the dead?} \citep{sheps_shall_1958}, a landmark paper that was published in the New England Journal of Medicine, and expanded upon with additional theoretical detail by the same author in Biometrics\citep{sheps_examination_1959}. Remarkably, Sheps' approach not only resolves the theoretical problems of closure and invariance, it also grounds choice of effect measure in biological mechanisms consistent with influential work from toxicology, psychology and philosophy. Variations of Sheps' insight have been repeatedly rediscovered over the last 65 years \citep{weinberg_applicability_1986, cheng_covariation_1997, bouckaert_sure_2001, huitfeldt_choice_2018, baker_new_2018}, but remain largely unknown among clinicians, epidemiologists and other practitioners of applied statistics.

In this commentary, we refer to the variant of the relative risk that is based on the probability of the outcome event as the standard risk ratio ($RR=\frac{Pr[Y^{a=1}=1]}{Pr[Y^{a=0}=1]}$)
and to the variant that is based on the complement of the outcome event as the survival ratio ($SR=\frac{Pr[Y^{a=1}=0]}{Pr[Y^{a=0}=0]}$).

\section{Sheps' argument}

Sheps' premise is stated very clearly in the introduction to her 1958 paper: ``When we wish to estimate the beneficial effect of a therapeutic agent, or the harmful effect of a factor such as exposure to radiation, we have the choice of a number of methods. If a new treatment is associated with lower mortality rates, we can estimate the beneficial effect from the absolute increase in the number of survivors, from the percentage increase in the living, or from the percentage decrease in the dead. Often, these different methods provide estimates of quite different magnitude. (...) However, in the assessment of beneficial or harmful effects respectively, there may be cogent reasons for choosing particular methods''.

To illustrate the setting of beneficial effects, she considers the Polio vaccine, and shows that, under an assumption of monotonicity (i.e. that the vaccine does not cause Polio in any individual), the relative decrease in probability of polio, which with counterfactual notation would be written as $\frac{Pr[Y^{a=0}=1]-Pr[Y^{a=1}=1]}{Pr[Y^{a=0}=1]}$, can be interpreted as reflecting the proportion in whom the vaccine prevents polio (i.e. the prevalence of the “preventive” response type, as described in \citet{greenland_identifiability_1986}) among those who would get polio if unvaccinated. In contrast, the relative increase in probability of not getting polio ($\frac{Pr[Y^{a=1}=0] -Pr[Y^{a=0}=0]}{Pr[Y^{a=0}=0]}$) has no interpretation in terms of prevalence of response types. The relative decrease in probability of the outcome event is equal to $1-RR$, modelling the relative decrease is therefore equivalent to modelling the standard risk ratio. 

Similarly, to illustrate the setting of harmful effects, she considers the effect of smoking on mortality, and shows that under a monotonicity assumption, the relative decrease in probability of survival ($\frac{Pr[Y^{a=0}=0]-Pr[Y^{a=1}=0]}{Pr[Y^{a=0}=0]}$) can be interpreted as the proportion who are harmed (i.e. the prevalence of the “causal” response type) among those who would have survived if they didn't smoke. In contrast, the relative increase in probability of death ($\frac{Pr[Y^{a=1}=1]-Pr[Y^{a=0}=1]}{Pr[Y^{a=0}=1]}$) has no interpretation in terms of prevalence of response types. The relative decrease in the probability of the complement of the outcome event is equal to 1-SR, modelling the relative increase is therefore equivalent to modelling the survival ratio. 

Analogous arguments will hold in many settings: Beneficial interventions act on potential deaths or failures, making the potential failures the natural denominator for measuring the effect (leading to a preference for modelling the standard risk ratio). In contrast, harmful interventions act on potential survivors or successes (leading to a preference for modelling the survival ratio).

\section{A simple causal model}

When measuring the effect of medications, a central consideration for choice of scale is stability: A good effect measure is one that is believed to take approximately the same value in different groups, given reasonable assumptions about biological mechanisms. In order to illustrate the theoretical advantages of Sheps' approach, we will consider a simplified, highly stylized causal model that leads to stability of her preferred variant of the relative risk. While this model is simple, it is consistent with influential work from a variety of academic traditions, including sufficient component cause (``causal pie'') models\citep{rothman_causes_1976, huitfeldt_shall_2022}. 

Consider (unmeasured) individual-level attributes that determine whether a person responds to treatment. We refer to these attributes as ``switches'': if no switch is present in a person, treatment has no effect on them, whereas when switches are present in a person, they combine to determine whether and how treatment affects the outcome. We will refer to combinations of switches as ``switch patterns''; there are four logically possible ways that such switch patterns can operate, which are shown in Table 1.

\begin{table}[!htbp]
\centering
\footnotesize
\caption{Switch patterns}
   \begin{tabular}{|m{2cm}|m{2cm}|m{2cm}|m{2cm}|m{2.5cm}|m{2cm}m{2.5cm}|}
\hline
 \textbf{Type} & \textbf{Function} &\textbf{Equivalent description of function} &\textbf{Direction of effect} & \textbf{Example} &\multicolumn{2}{m{4.5cm}|}{\textbf{Implications when this switch pattern is solely responsible for the effect, and independent of baseline risk\newline }} \\
& & & & & Expression for prevalence of switch & Effect measure stability\\
\end{tabular}
\begin{tabular}{|m{2cm}|m{2cm}|m{2cm}|m{2cm}|m{2.5cm}|m{2cm}|m{2.5cm}|}
\hline
        ``Sufficient-causal''& Treatment is a sufficient cause of the outcome & Absence of treatment is a necessary cause of absence of the outcome&Treatment increases risk & Genetic variant such that carriers get an allergic reaction if they take the drug & $1-\frac{Pr[Y^{a=1}=0]}{Pr[Y^{a=0}=0]}$ & $SR$ will be stable between groups with similar prevalence of the switch pattern\\
        \hline
        ``Necessary-preventive''&Treatment is a necessary cause of absence of the outcome &Absence of treatment is a sufficient cause of the outcome&Treatment decreases risk&Bacterial strain such that in patients who are infected with that strain, the outcome event will always occur unless they take a specific drug &$1-\frac{Pr[Y^{a=0}=0]}{Pr[Y^{a=1}=0]}$&$\frac{1}{SR}$ will be stable between groups with similar prevalence of the switch pattern  \\
        \hline
        ``Sufficient-preventive''&Treatment is a sufficient cause of absence of the outcome &Absence of treatment is a necessary cause of the outcome&Treatment decreases risk&Bacterial strain such that in patients who are infected with that strain, the drug always prevents the outcome &$1-\frac{Pr[Y^{a=1}=1]}{Pr[Y^{a=0}=1]}$&$RR$ will be stable between groups with similar prevalence of the switch pattern  \\
        \hline
        ``Sufficient-causal''& Treatment is a necessary cause of the outcome & Absence of treatment is a sufficient cause of absence of the outcome&Treatment increases risk & Genetic variant such that no allergic reactions can occur in the carrier unless they take the drug& $1-\frac{Pr[Y^{a=0}=1]}{Pr[Y^{a=1}=1]}$ & $\frac{1}{RR}$ will be stable between groups with similar prevalence of the switch pattern\\
      \bottomrule
\end{tabular}
\end{table}

If the same switch pattern type accounts for treatment response in all individuals and the switch pattern is independent of baseline risk, then there will exist some variant of the relative risk whose value, in any group, reflects the prevalence of the switch pattern, as shown in table 1. When this is the case, predictors of switch pattern prevalence are effect modifiers on that scale\citep{huitfeldt_effect_2019}.

Reality is unfortunately rarely that simple: Any reasonable epidemiologist would certainly be wary about relying on an assumption that only one type of switch pattern accounts for the effect of treatment. However, in some cases, one type of switch is predominantly responsible for the effect of treatment, and in these cases, choosing a model based on the associated variant of the relative risk may allow heterogeneity to be understood in terms of biologically interpretable deviation from the most applicable model. 

In many specific cases, it is possible to show that the sufficient-causal pattern is more consistent with how biology is typically understood than the necessary-causal pattern (leading to a preference for the survival ratio for exposures that increase incidence, in accordance with Sheps' recommendation), and that the sufficient-preventive pattern is more consistent with biology than the necessary-preventive pattern (leading to a preference for the risk ratio for exposures that reduce incidence, again in accordance with Sheps' recommendation). For example, in the setting of a vaccine, a sufficient-preventive pattern may reflect the fact that the vaccine is sufficient to prevent the disease in those who are exposed to a particular susceptible strain of the virus, and has little effect in those who are exposed to a non-susceptible strain. In contrast, it would be challenging to describe a biological phenomenon that can be represented as a necessary-preventive pattern. Similar arguments can be made in many clinically relevant applications, almost always leading to a preference for the variant of the relative risk suggested by Sheps. Table 1 provides some additional such examples.

But if it is true, as a feature of nature, that some switch patterns are much more common than others, this screams out for an explanation: How did biology end up being so asymmetric? In one potential attempt to account for this asymmetry, it may be useful to invoke evolutionary theory. In an environment where everybody is unexposed, the sufficient-causal and the sufficient-preventive pattern types are both inert, whereas the necessary-causal and the necessary-preventive pattern types strongly determine the outcome. Similarly, in an environment where everybody is exposed, the necessary-causal and necessary-preventive pattern types are both inert, whereas the sufficient-causal and sufficient-preventive types strongly determine the outcome.  

The switch pattern types that lead to stability of Sheps' preferred variant of the relative risk, are exactly those pattern types that would have been inert in evolutionary history, as modern medical treatments were almost always unavailable to our ancestors. This hints at a possible explanation for nature's asymmetry: Only those switch pattern types that lead to Sheps' conclusions would have been inert in evolutionary history, and could plausibly remain in our gene pool as determinants of variation in treatment response.

\section{A brief history of related ideas}
While Sheps was the first to suggest this line of reasoning in medical statistics, her ideas were anticipated by researchers working in toxicology and entomology. In a very short paper from 1925, W.S. Abbott, an entomologist working at the US Department of Agriculture, proposed measuring the effect of insect sprays using what is now known as Abbott’s Formula\citep{abbott_method_1925}. This formula, which is still used by entomologists today, is equivalent to Sheps’ suggestion for the case where the intervention increases risk of the outcome. 

Abbott did not consider the case where exposure reduces risk of the outcome, which is not surprising because he studied insecticides. That consideration was left to another another entomologist, albeit one who played a significant role in the early modern history of statistics: many readers will recognize C.I. Bliss as the originator of the Probit Model. In 1939, he extended Abbott’s formula to the setting where exposure reduces incidence, preserving its equivalence with Sheps' approach. To do so, he developed the Joint Independent Action model\citep{bliss_toxicity_1939} , a simple mechanism for drug effect which has become central to how toxicologists think about interactions between poisons. 

Attempts to introduce these approaches in epidemiology have been made by several researchers whose research takes place at the intersection between toxicology and epidemiology. For example, Clarice Weinberg showed that when a drug acts according to Abbott’s model for the mechanism of action, generalized linear models should use a log link when considering outcomes whose incidence are reduced by the drug, and a complementary log link when considering outcomes whose incidence are reduced by the drug\citep{weinberg_applicability_1986, weinberg_can_2007, weinberg_interaction_2012}. This modelling strategy leads to estimating main effects that are consistent with Sheps' recommendations. However, it is rarely used in practice, possibly because it would require data analysts to know the direction of the effect before specifying the model.

A solution to this problem was provided by \citet{van_der_laan_estimation_2007}, who proposed the switch relative risk, an effect parameter which automatically selects a variant of the relative risk in accordance with Sheps’ principles. Independently, \citet{baker_new_2018} proposed the generalized relative risk reduction (GRRR), a closely related effect measure which encodes all the relevant information on the interval [-1,1]. 

Models that imply stability of Sheps' preferred variant of the relative risk are also influential in the philosophy and psychology literatures, in the form of Patricia Cheng’s Power PC theory\citep{cheng_covariation_1997}. It has been argued on theoretical and empirical grounds that human decision makers use (and should use) these models to carry causal information from one context to another\citep{liljeholm_when_2007}, i.e. for purposes of generalizability. Clark Glymour, who is a leading philosopher of science, referred to the Power PC model as a "brilliant piece of mathematical metaphysics”\citep{glymour_minds_2001}.

\section{Conclusions}

Among the scales that can be used to measure the effects of medical interventions, Sheps' approach is unique in that stability of her preferred variant of the relative risk may result from well-understood and plausible biological mechanisms. The underlying ideas have been independently rediscovered multiple times, by researchers from academic traditions as diverse as entomology, toxicology, psychology, philosophy, computer science and epidemiology. The diversity of thinkers who have arrived independently at the same conclusions, suggests that Sheps’ ideas represent an attractor in idea space, and that they will continue to be rediscovered as long as her groundbreaking work remains relatively unknown and unused in practice.

\bibliographystyle{rss}
\bibliography{references}

\end{document}